\documentclass[aip,preprint,endfloats]{revtex4-1}
\usepackage{graphicx}
\usepackage{dcolumn}
\usepackage{bm}
\usepackage{color}

\begin{document}

\preprint{Draft}

\title{Dependence of nonlocal Gilbert damping on the ferromagnetic layer type in FM/Cu/Pt heterostructures}

\author{ A. Ghosh, J.F. Sierra, S. Auffret, U. Ebels}
\affiliation{SPINTEC, UMR(8191) CEA / CNRS / UJF / Grenoble INP ; INAC, 17 rue des Martyrs, 38054 Grenoble Cedex, France}
\author{W.E. Bailey}
\affiliation{Dept. of Applied Physics \& Applied Mathematics, Columbia University, New York NY 10027, USA}

\date{\today}
\begin{abstract}
We have measured the size effect in nonlocal Gilbert relaxation rate in FM(t$_{FM}$) / Cu (5nm) [/ Pt (2nm)] / Al(2nm) heterostructures,  FM = \{ Ni$_{81}$Fe$_{19}$, Co$_{60}$Fe$_{20}$B$_{20}$,  pure Co\}.  Common behavior is observed for three FM layers, where the additional relaxation obeys both a strict inverse power law dependence $\Delta G =K \:t^{n}$, $n=-\textrm{1.04}\pm\textrm{0.06}$ and a similar magnitude $K=\textrm{224}\pm\textrm{40 Mhz}\cdot\textrm{nm}$.  As the tested FM layers span an order of magnitude in spin diffusion length $\lambda_{SDL}$, the results are in support of spin diffusion, rather than nonlocal resistivity, as the origin of the effect.
\end{abstract}

\maketitle

The primary materials parameter which describes the temporal response of magnetization $\mathbf{M}$ to applied fields $\mathbf{H}$ is the Gilbert damping parameter $\alpha$, or relaxation rate $G=|\gamma| M_s\alpha$.  Understanding of the Gilbert relaxation, particularly in structures of reduced dimension, is an essential question for optimizing the high speed / Ghz response of nanoscale magnetic devices.

Experiments over the last decade have established that the Gilbert relaxation of ferromagnetic ultrathin films exhibits a size effect, some component of which is nonlocal.  Both $\alpha(t_{FM})=\alpha_{0}+\alpha^{'}(t_{FM})$ and $G(t_{FM})=G_{0}+G^{'}(t_{FM})$ increase severalfold with decreasing FM film thickness $t_{FM}$, from near-bulk values $\alpha_{0}, G_{0}$ for $t_{FM}\gtrsim\textrm{20 nm}$.  Moreover, the damping size effect can have a nonlocal contribution responsive to layers or scattering centers removed, through a nonmagnetic (NM) layer, from the precessing FM.  Contributed Gilbert relaxation has been seen from other FM layers\cite{heinrich-urban-gilbert} as well as from heavy-element scattering layers such as Pt.\cite{miz3}

The nonlocal damping size effect is strongly reminiscent of the electrical resistivity in ferromagnetic ultrathin films.  Electrical resistivity $\rho$ is size-dependent by a similar factor over a similar range of $t_{FM}$; the resistivity $\rho(t_{FM})$ is similarly nonlocal, dependent upon layers not in direct contact.\cite{dienyfundamental,butlerPRL,bailey-tsymbal}.  It is {\it prima facie} plausible that the nonlocal damping and nonlocal electrical resistivity share a common origin in momentum scattering (with relaxation time $\tau_{M}$) by overlayers.  If the nonlocal damping arises from nonlocal scattering $\tau_{M}^{-1}$, however, there should be a marked dependence upon FM layer type.   Damping in materials with short spin diffusion length $\lambda_{SDL}$ is thought to be proportional to $\tau_M^{-1}$ (ref.\cite{kambersky-microscopic}); the claim for "resistivity-like" damping has been made explicitly for Ni$_{81}$Fe$_{19}$ by Ingvarsson\cite{ingvarsson-scattering} et al.  For FM with a long $\lambda_{SDL}$, on the other hand, relaxation $G$ is either nearly constant with temperature or "conductivity-like," scaling as $\tau_{M}$.

Interpretation of the nonlocal damping size effect has centered instead on a spin current model\cite{silsbeePRB79} advanced by Tserkovnyak et al\cite{tserkovPRL02sp}.  An explicit prediction of this model is that the magnitude of the nonlocal Gilbert relaxation rate $\Delta G$ is only weakly dependent upon the FM layer type.  The effect has been calculated\cite{tserk-rmp} as \begin{equation}\Delta G=|\gamma|^{2}\hbar/ 4\pi\left({g_{eff}^{\uparrow\downarrow}/
S}\right) t_{FM}^{-1}\label{speq}\end{equation}, where the effective spin mixing conductance $g_{eff}^{\uparrow\downarrow}/S$ is given in units of channels per area.  {\it Ab-initio} calculations predict a very weak materials dependence for the interfacial parameters $g^{\uparrow\downarrow}/S$, with $\pm\textrm{10}$\% difference in systems as different as Fe/Au and Co/Cu, and negligible dependence on interfacial mixing.\cite{bauerPRB05}

Individual measurements exist of the spin mixing conductance, through the damping, in FM systems Ni$_{81}$Fe$_{19}$\cite{miz1,*miz2}, Co\cite{beaujourJAP06,*beaujourPRB06}, and CoFeB\cite{mewesJPD08}.  However, these experiments do not share a common methodology, which makes a numerical comparison of the results problematic, especially given that Gilbert damping estimates are to some extent model-dependent\cite{mcmichael-semiclassical}.  In our experiments, we have taken care to isolate the nonlocal damping contribution due to Pt overlayers only, controlling for growth effects, interfacial intermixing, and inhomogeneous losses.  The only variable in our comparison of nonlocal damping $\Delta G(t_{FM})$, to the extent possible, has been the identity of the FM layer.

Gilbert damping $\alpha$ has been measured through ferromagnetic resonance (FMR) from $\omega/2\pi=\textrm{2-24 Ghz}$ using a broadband coplanar waveguide (CPW) with broad center conductor width $w=\textrm{400}\mu m$, using field modulation and lock-in detection of the transmitted signal to enhance sensitivity.  The Gilbert damping has been separated from inhomogeneous broadening in the films measured using the well-known relation $\Delta H_{pp}(\omega)=\Delta H_{0}+\left(2/\sqrt{3}\right)\alpha\omega/|\gamma|$.  We have fit spectra to Lorenzian derivatives with Dysonian components at each frequency, for each film, to extract the linewidth $\Delta H_{pp}$ and resonance field $H_{res}$; $\alpha$ has been extracted using linear fits to $\Delta H(\omega)$.

For the films, six series of heterostructures were deposited of the form Si/ SiO$_{2}$/ X/ FM($t_{FM}$)/ Cu(3nm)$\mathbf{[/Pt(3nm)]}$/ Al(3nm), FM = \{ Ni$_{81}$Fe$_{19}$ ("Py"), Co$_{60}$Fe$_{20}$B$_{20}$ ("CoFeB"),  pure Co\}, and  $t_{FM}=\textrm{2.5, 3.5, 6.0, 10.0, 17.5, 30.0 nm}$, for 36 heterostructures included in the study.  For each ferromagnetic layer type $FM$, one thickness series $t_{FM}$ was deposited with the Pt overlayer and one thickness series $t_{FM}$ was deposited without the Pt overlayer.  This makes it possible to record the additional damping $\Delta\alpha(t_{FM})$ introduced by the Pt overlayer alone, independent of size effects present in the FM/Cu layers deposited below. In the case of pure Co, a X=Ta(5nm)/Cu(5nm) underlayer was necessary to stabilize low-linewidth films, otherwise, depositions were carried out directly upon the {\it in-situ} ion-cleaned substrate.

\begin{figure}
\includegraphics{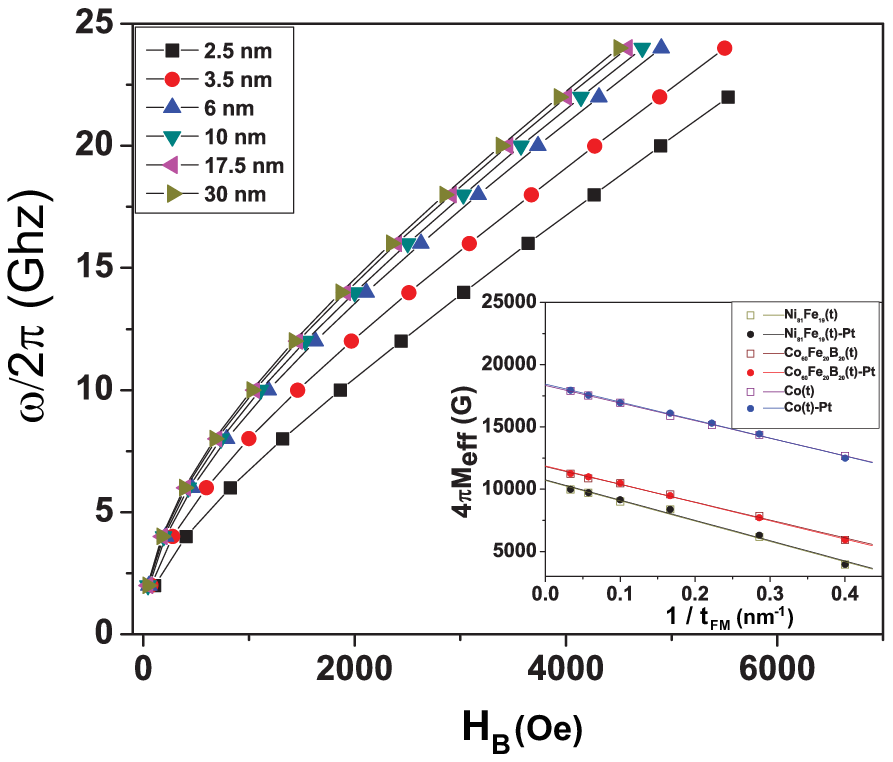}
\centering \caption{Fields for resonance $\omega(H_{B})$ for in-plane FMR, FM=Ni$_{81}$Fe$_{19}$, $\textrm{2.5 nm}\leq t_{FM}\leq\textrm{30.0 nm}$; solid lines are Kittel fits.  {\it Inset:} 4$\pi$ M$_{s}^{eff}$ for all three FM/Cu, with and without Pt overlayers.}\label{fig1}\end{figure}

Field-for-resonance data are presented in Figure \ref{fig1}.  The main panel shows $\omega(H_{B}^{\parallel})$ data for Ni$_{81}$Fe$_{19}(t_{FM})$.  Note that there is a size effect in $\omega(H_{B}^{\parallel})$: the thinner films have a substantially lower resonance frequency.  For $t_{FM}=\textrm{2.5 nm}$, the resonance frequency is depressed by $\sim$5 Ghz from $\sim$20 Ghz resonance $H_{B}\simeq\textrm{4 kOe}$.  The behavior is fitted through the Kittel relation (lines) $\omega(H_{B}^{\parallel})=|\gamma|\sqrt{\left(H_{B}^{\parallel}+H_{K}\right)\left(4\pi M_s^{eff}+H_{B}^{\parallel}+H_{K}\right)}$, and the inset shows a summary of extracted $4\pi M_s^{eff}(t_{FM})$ data for the three different FM layers. Samples with (open symbols) and without (closed symbols) Pt overlayers show negligible differences. Linear fits according to $4\pi M_s^{eff}(t_{FM})=4\pi M_s-\left(2 K_{s}/M_{s}\right)t_{FM}^{-1}$ allow the extraction of bulk magnetization $4\pi M_s$ and surface anisotropy $K_s$; we find $4\pi M_s^{Py}=\textrm{10.7 kG}$, $4\pi M_s^{CoFeB}=\textrm{11.8 kG}$, $4\pi M_s^{Co}=\textrm{18.3 kG}$, and $K_s^{Py}=\textrm{0.69 erg/cm}^2$, $K_s^{CoFeB}=\textrm{0.69 erg/cm}^2$, $K_s^{Co}=\textrm{1.04 erg/cm}^2$.  The value of $g_{L}/2=|\gamma|/(e/mc)$, $|\gamma|=2\pi\cdot(\textrm{2.799 Mhz/Oe})\cdot\left(g_{L}/2\right)$ is found from the Kittel fits subject to this choice, yielding $g_{L}^{Py}=\textrm{2.09}$, $g_{L}^{CoFeB}=\textrm{2.07}$, $g_{L}^{Co}=\textrm{2.15}$.  The $4\pi M_{s}$ and $g_{L}$ values, taken to be size-independent, are in good agreement with bulk values.

\begin{figure}
\includegraphics{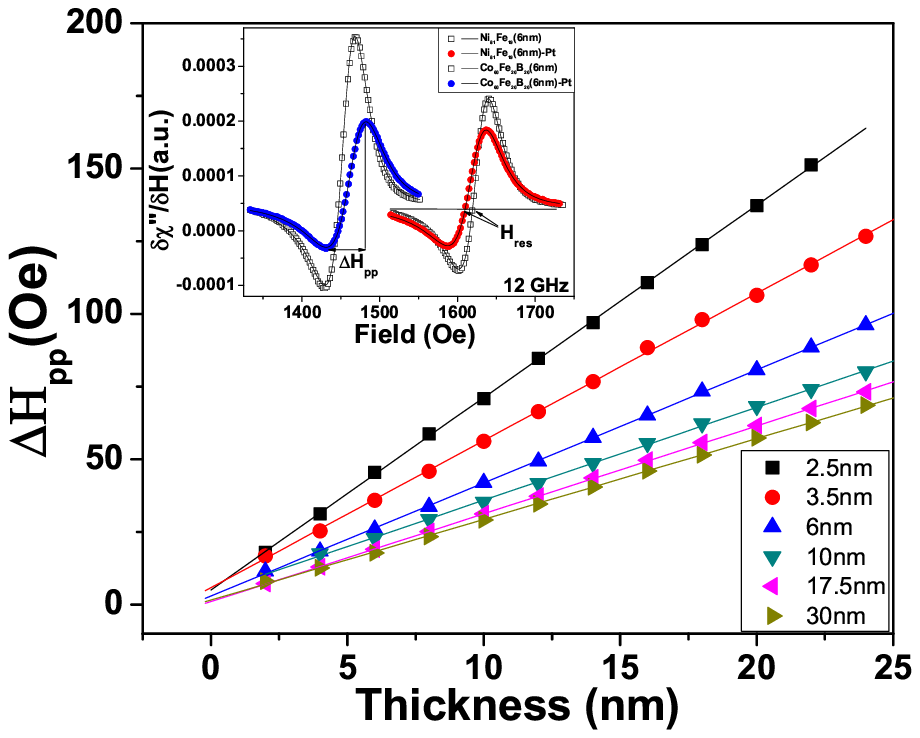}
\centering \caption{Frequency-dependent peak-to-peak FMR linewidth $\Delta H_{pp}(\omega)$ for FM=Ni$_{81}$Fe$_{19}$, $t_{FM}$ as noted, films with Pt overlayers.  {\it Inset:} lineshapes and fits for films with and without Pt, FM=Ni$_{81}$Fe$_{19}$, CoFeB.}\label{fig2}\end{figure}

FMR linewidth as a function of frequency $\Delta H_{pp}(\omega)$ is plotted in Figure \ref{fig2}.  The data for Py show a near-proportionality, with negligble inhomogeneous component $\Delta H_{0}\leq\textrm{4 Oe}$ even for the the thinnest layers, facilitating the extraction of intrinsic damping parameter $\alpha$.  The size effect in in $\alpha(t_{FM})$ accounts for an increase by a factor of $\sim$ 3, from $\alpha_{0}^{Py}=\textrm{0.0067} \:(G_{0}^{Py}=\textrm{105 Mhz})$ for the thickest films ($t_{FM}=\textrm{30.0 nm}$) to $\alpha=\textrm{0.021}$  for the thinnest films ($t_{FM}=\textrm{2.5 nm}$).  The inset shows the line shapes for films with and without Pt, illustrating the broadening without significant frequency shift or significant change in peak asymmetry.

A similar analysis has been carried through for CoFeB and Co (not pictured).  Larger inhomogeneous linewidths are observed for pure Co, but homogeneous linewidth still exceeds inhomogeneous linewidth by a factor of three over the frequency range studied, and inhomogeneous linewidths agree within experimental error for the thinnest films with and without Pt overlayers.  We extract for these films $\alpha_{0}^{CoFeB}=\textrm{0.0065} \: (G_{0}^{CoFeB}=\textrm{111 Mhz})$ and $\alpha_{0}^{Co}=\textrm{0.0085} \: (G_{0}^{Co}=\textrm{234 Mhz})$.  The latter value is in very good agreement with the average of easy- and hard-axis values for epitaxial FCC Co films measured up to 90 Ghz, $G_{0}^{Co}=\textrm{225 Mhz.}$\cite{Schreiber95SSC}

\begin{figure}
\includegraphics{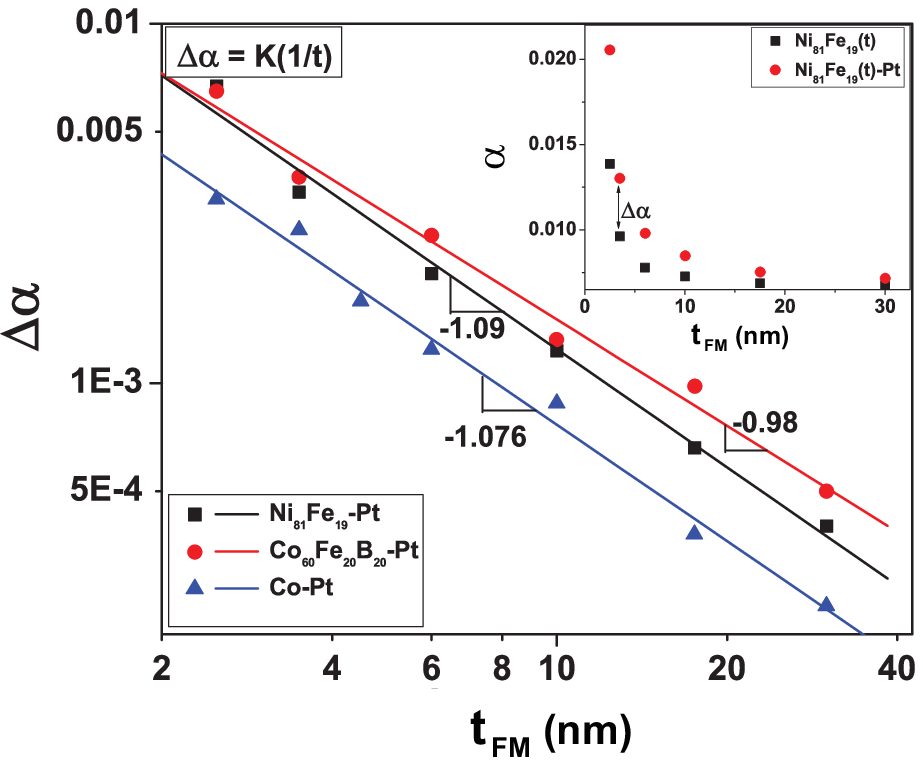}
\centering \caption{{\it Inset: } $\alpha_{\textrm{no Pt}}(t_{FM})$ and $\alpha_{\textrm{Pt}}$ for Py, after linear fits to data in Figure \ref{fig2}. Main panel: $\Delta\alpha(t_{FM})=\alpha_{\textrm{Pt}}(t_{FM})-\alpha_{\textrm{no Pt}}(t_{FM})$ for Py, CoFeB, and Co.  The slopes express the power law exponent $n=\textrm{-1.04}\pm\textrm{0.06}.$}\label{fig3}\end{figure}

We isolate the effect of Pt overlayers on the damping size effect in Figure \ref{fig3}.  Values of $\alpha$ have been fitted for each deposited heterostructure: each FM type, at each $t_{FM}$, for films with and without Pt overlayers.  We take the difference $\Delta\alpha(t_{FM})$ for identical FM($t_{FM}$)/Cu(5nm)/Al(3nm) depositions with and without the insertion of Pt(3nm) after the Cu deposition.  Data, as shown on the logarithmic plot in the main panel, are found to obey a power law $\Delta\alpha(t_{FM})=K t^{n}$, with $n=\textrm{-1.04}\pm\textrm{0.06}$.  This is excellent agreement with an inverse thickness dependence $\Delta \alpha(t_{FM})=K_{FM}/t_{FM}$, where the prefactor clearly depends on the FM layer, highest for Py and lowest for Co.  Note that efforts to extract $\Delta\alpha(t_{FM})=K t^{n}$ without the FM($t_{FM}$)/Cu baselines would meet with significant errors; numerical fits to $\alpha(t_{FM})=K t_{FM}\:^{n}$ for the FM($t_{FM}$)/Cu/Pt structures yield exponents $n\simeq\textrm{1.4}$.

\begin{figure}
\includegraphics{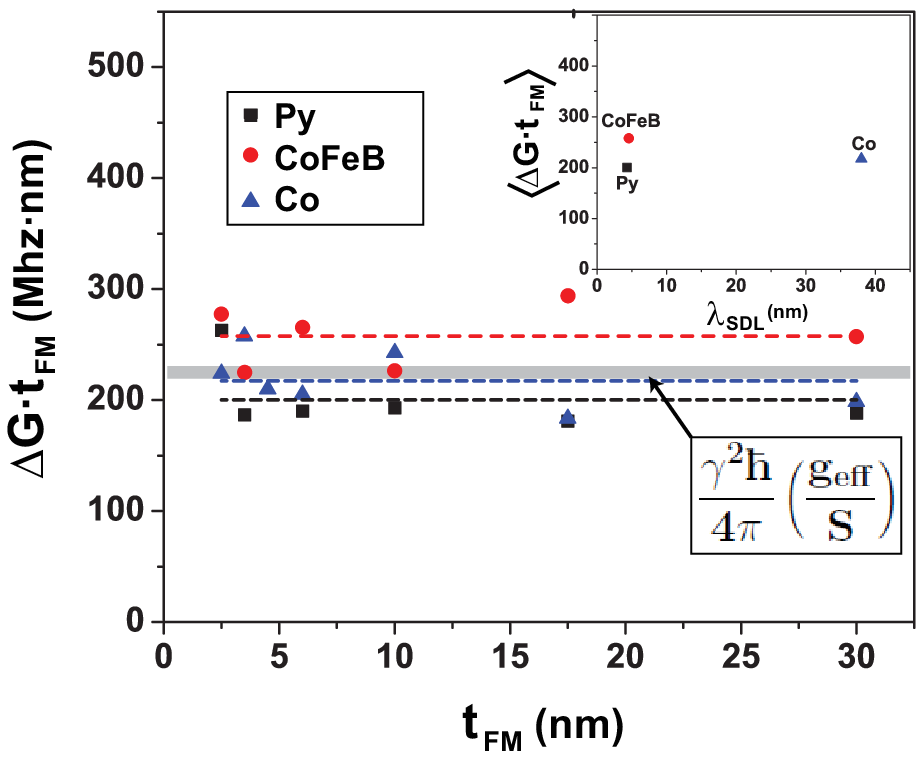}
\centering \caption{The additional nonlocal relaxation due to Pt overlayers, expressed as a Gilbert relaxation rate - thickness product $\Delta G\cdot t_{FM}$ for Py, CoFeB, and Co.  {\it Inset:} dependence of $\Delta G\cdot t_{FM}$ on spin diffusion length $\lambda_{SDL}$ as tabulated in \cite{bassJPCM07,*prattAPL08}.
}\label{fig4}\end{figure}

Expressing now the additional Gilbert relaxation as $\Delta G(t_{FM})=|\gamma|M_{s}\Delta\alpha(t_{FM})=|\gamma^{FM}|M_{s}^{FM}K_{FM}/t_{FM}$, we plot $\Delta G\cdot t_{FM}$ in Figure \ref{fig4}.  We find $\Delta G\cdot t_{Py}=\textrm{192}\pm\textrm{40 Mhz}$, $\Delta G\cdot t_{CoFeB}=\textrm{265}\pm\textrm{40 Mhz}$, and $\Delta G\cdot t_{Co}=\textrm{216}\pm\textrm{40 Mhz}$.  The similarity of values for $\Delta G\cdot t_{FM}$ is in good agreement with predictions of the spin pumping model in Equation \ref{speq}, given that interfacial spin mixing parameters are nearly equal in different systems.

The similarity of the $\Delta G\cdot t_{FM}$ values for the different FM layers is, however, at odds with expectations from the "resistivity-like" mechanism.  In Figure \ref{fig4}, {\it inset}, we show the dependence of $\Delta G \cdot t_{FM}$ upon the tabulated $\lambda_{SDL}$ of these layers from Ref \cite{bassJPCM07,*prattAPL08}.  It can be seen that $\lambda_{SDL}^{Co}$ is roughly an order of magnitude longer than it is for the other two FM layers, Py and CoFeB, but the contribution of Pt overlayers to damping is very close to their average.  Since under the resistivity mechanism, only Py and CoFeB should be susceptible to a resistivity contribution in $\Delta\alpha(t_{FM})$, the results imply that the contribution of Pt to the nonlocal damping size effect has a separate origin.

Finally, we compare the magnitude of the nonlocal damping size effect with that predicted by the spin pumping model in Ref. \cite{tserk-rmp}.  According to $\Delta G\cdot t_{FM}=|\gamma|^{2}\hbar/ 4\pi=\textrm{25.69
Mhz}\cdot\textrm{nm}^3(g_{L}/2)^2\left({g_{eff}^{\uparrow\downarrow}/
S}\right)$, our experimental $\Delta G\cdot t_{FM}$ and $g_{L}$ data yield effective spin mixing conductances $g_{eff}^{\uparrow\downarrow}/
S\left[\textrm{\small{Py/Cu/Pt}}\right]=\textrm{6.8 nm}^{-2}$, $g_{eff}^{\uparrow\downarrow}/
S\left[\textrm{\small{Co/Cu/Pt}}\right]=\textrm{7.3 nm}^{-2}$, and $g_{eff}^{\uparrow\downarrow}/
S\left[\textrm{\small{CoFeB/Cu/Pt}}\right]=\textrm{9.6 nm}^{-2}$.  The Sharvin-corrected form, in the realistic limit of $\lambda_{SDL}^{N}\gg t_{N}$\cite{bauerPRB05} is $
({g^{\uparrow\downarrow}_{eff}/ S})^{-1} = ({g^{\uparrow\downarrow}_{F/N}/ S})^{-1} -{1\over 2}({g^{\uparrow\downarrow}_{N,S}/ S})^{-1} + {2 e^2}h^{-1}\rho\:t_{N} + ({\tilde{g}^{\uparrow\downarrow}_{N_{1}/N_{2}}/ S})^{-1}$.  Using conductances 14.1nm$^{-2}$ (Co/Cu), 15.0nm$^{-2}$ (Cu), 211nm$^{-2}$ (bulk $\rho_{Cu}$, $t_{N}=\textrm{3nm}$), 35 nm$^{-2}$ (Cu/Pt) would predict a theoretical $g_{eff,th.}^{\uparrow\downarrow}/
S\left[\textrm{\small{Co/Cu/Pt}}\right]=\textrm{14.1 nm}^{-2}$.  Reconciling theory and experiment would require an order of magnitude larger $\rho_{Cu}\simeq\textrm{20}\mu\Omega\cdot\textrm{cm}$, likely not physical.

To summarize, a common methodology, controlling for damping size effects and intermixing in single films, has allowed us to compare the nonlocal damping size effect in different FM layers.  We observe, for Cu/Pt overlayers, the same power law in thickness $t^{-\textrm{1.04}\pm\textrm{0.06}}$, the same materials independence, but roughly half the magnitude that predicted by the spin pumping theory of Tserkovnyak\cite{tserk-rmp}.  The rough independence on FM spin diffusion length, shown here for the first time, argues against a resistivity-based interpretation for the effect.

We would like to acknowledge the US NSF-ECCS-0925829, the Bourse Accueil Pro n$^\circ$ 2715 of the Rh\^{o}ne-Alpes Region, the French National Research Agency (ANR) Grant ANR-09-NANO-037, and the FP7-People-2009-IEF program no 252067.

%


\begin{thebibliography}{20}%
\makeatletter
\providecommand \@ifxundefined [1]{%
 \@ifx{#1\undefined}
}%
\providecommand \@ifnum [1]{%
 \ifnum #1\expandafter \@firstoftwo
 \else \expandafter \@secondoftwo
 \fi
}%
\providecommand \@ifx [1]{%
 \ifx #1\expandafter \@firstoftwo
 \else \expandafter \@secondoftwo
 \fi
}%
\providecommand \natexlab [1]{#1}%
\providecommand \enquote  [1]{``#1''}%
\providecommand \bibnamefont  [1]{#1}%
\providecommand \bibfnamefont [1]{#1}%
\providecommand \citenamefont [1]{#1}%
\providecommand \href@noop [0]{\@secondoftwo}%
\providecommand \href [0]{\begingroup \@sanitize@url \@href}%
\providecommand \@href[1]{\@@startlink{#1}\@@href}%
\providecommand \@@href[1]{\endgroup#1\@@endlink}%
\providecommand \@sanitize@url [0]{\catcode `\\12\catcode `\$12\catcode
  `\&12\catcode `\#12\catcode `\^12\catcode `\_12\catcode `\%12\relax}%
\providecommand \@@startlink[1]{}%
\providecommand \@@endlink[0]{}%
\providecommand \url  [0]{\begingroup\@sanitize@url \@url }%
\providecommand \@url [1]{\endgroup\@href {#1}{\urlprefix }}%
\providecommand \urlprefix  [0]{URL }%
\providecommand \Eprint [0]{\href }%
\providecommand \doibase [0]{http://dx.doi.org/}%
\providecommand \selectlanguage [0]{\@gobble}%
\providecommand \bibinfo  [0]{\@secondoftwo}%
\providecommand \bibfield  [0]{\@secondoftwo}%
\providecommand \translation [1]{[#1]}%
\providecommand \BibitemOpen [0]{}%
\providecommand \bibitemStop [0]{}%
\providecommand \bibitemNoStop [0]{.\EOS\space}%
\providecommand \EOS [0]{\spacefactor3000\relax}%
\providecommand \BibitemShut  [1]{\csname bibitem#1\endcsname}%
\let\auto@bib@innerbib\@empty
\bibitem [{\citenamefont {Urban}, \citenamefont {Woltersdorf},\ and\
  \citenamefont {Heinrich}(2001)}]{heinrich-urban-gilbert}%
  \BibitemOpen
  \bibfield  {author} {\bibinfo {author} {\bibfnamefont {R.}~\bibnamefont
  {Urban}}, \bibinfo {author} {\bibfnamefont {G.}~\bibnamefont {Woltersdorf}},
  \ and\ \bibinfo {author} {\bibfnamefont {B.}~\bibnamefont {Heinrich}},\
  }\bibfield  {title} {\enquote {\bibinfo {title} {Gilbert damping in single
  and multilayer ultrathin films: role of interfaces in nonlocal spin
  dynamics},}\ }\href@noop {} {\bibfield  {journal} {\bibinfo  {journal}
  {Physical Review Letters}\ }\textbf {\bibinfo {volume} {87}},\ \bibinfo
  {pages} {217204--7} (\bibinfo {year} {2001})}\BibitemShut {NoStop}%
\bibitem [{\citenamefont {Mizukami}, \citenamefont {Ando},\ and\ \citenamefont
  {Miyazaki}(2002{\natexlab{a}})}]{miz3}%
  \BibitemOpen
  \bibfield  {author} {\bibinfo {author} {\bibfnamefont {S.}~\bibnamefont
  {Mizukami}}, \bibinfo {author} {\bibfnamefont {Y.}~\bibnamefont {Ando}}, \
  and\ \bibinfo {author} {\bibfnamefont {T.}~\bibnamefont {Miyazaki}},\
  }\bibfield  {title} {\enquote {\bibinfo {title} {Effect of spin diffusion on
  {G}ilbert damping for a very thin permalloy layer in {C}u/permalloy/{C}u/{P}t
  films},}\ }\href {\doibase 10.1103/PhysRevB.66.104413} {\bibfield  {journal}
  {\bibinfo  {journal} {Phys. Rev. B}\ }\textbf {\bibinfo {volume} {66}},\
  \bibinfo {pages} {104413} (\bibinfo {year} {2002}{\natexlab{a}})}\BibitemShut
  {NoStop}%
\bibitem [{\citenamefont {Dieny}\ \emph {et~al.}(1992)\citenamefont {Dieny},
  \citenamefont {Nozieres}, \citenamefont {Speriosu}, \citenamefont {Gurney},\
  and\ \citenamefont {Wilhoit}}]{dienyfundamental}%
  \BibitemOpen
  \bibfield  {author} {\bibinfo {author} {\bibfnamefont {B.}~\bibnamefont
  {Dieny}}, \bibinfo {author} {\bibfnamefont {J.}~\bibnamefont {Nozieres}},
  \bibinfo {author} {\bibfnamefont {V.}~\bibnamefont {Speriosu}}, \bibinfo
  {author} {\bibfnamefont {B.}~\bibnamefont {Gurney}}, \ and\ \bibinfo {author}
  {\bibfnamefont {D.}~\bibnamefont {Wilhoit}},\ }\bibfield  {title} {\enquote
  {\bibinfo {title} {Change in conductance is the fundamental measure of
  spin-valve magnetoresistance},}\ }\href@noop {} {\bibfield  {journal}
  {\bibinfo  {journal} {Applied Physics Letters}\ }\textbf {\bibinfo {volume}
  {61}},\ \bibinfo {pages} {2111--3} (\bibinfo {year} {1992})}\BibitemShut
  {NoStop}%
\bibitem [{\citenamefont {Butler}\ \emph {et~al.}(1996)\citenamefont {Butler},
  \citenamefont {Zhang}, \citenamefont {Nicholson}, \citenamefont
  {Schulthess},\ and\ \citenamefont {MacLaren}}]{butlerPRL}%
  \BibitemOpen
  \bibfield  {author} {\bibinfo {author} {\bibfnamefont {W.~H.}\ \bibnamefont
  {Butler}}, \bibinfo {author} {\bibfnamefont {X.~G.}\ \bibnamefont {Zhang}},
  \bibinfo {author} {\bibfnamefont {D.~M.~C.}\ \bibnamefont {Nicholson}},
  \bibinfo {author} {\bibfnamefont {T.~C.}\ \bibnamefont {Schulthess}}, \ and\
  \bibinfo {author} {\bibfnamefont {J.~M.}\ \bibnamefont {MacLaren}},\
  }\bibfield  {title} {\enquote {\bibinfo {title} {Giant magnetoresistance from
  an electron waveguide effect in cobalt-copper multilayers},}\ }\href@noop {}
  {\bibfield  {journal} {\bibinfo  {journal} {Physical Review Letters}\
  }\textbf {\bibinfo {volume} {76}},\ \bibinfo {pages} {3216--19} (\bibinfo
  {year} {1996})}\BibitemShut {NoStop}%
\bibitem [{\citenamefont {Bailey}, \citenamefont {Wang},\ and\ \citenamefont
  {Tsymbal}(2000)}]{bailey-tsymbal}%
  \BibitemOpen
  \bibfield  {author} {\bibinfo {author} {\bibfnamefont {W.~E.}\ \bibnamefont
  {Bailey}}, \bibinfo {author} {\bibfnamefont {S.~X.}\ \bibnamefont {Wang}}, \
  and\ \bibinfo {author} {\bibfnamefont {E.~Y.}\ \bibnamefont {Tsymbal}},\
  }\bibfield  {title} {\enquote {\bibinfo {title} {Electronic scattering from
  {C}o/{C}u interfaces: In situ measurement and comparison with theory},}\
  }\href {\doibase 10.1103/PhysRevB.61.1330} {\bibfield  {journal} {\bibinfo
  {journal} {Phys. Rev. B}\ }\textbf {\bibinfo {volume} {61}},\ \bibinfo
  {pages} {1330--1335} (\bibinfo {year} {2000})}\BibitemShut {NoStop}%
\bibitem [{\citenamefont {Kambersk{\'y}}(1970)}]{kambersky-microscopic}%
  \BibitemOpen
  \bibfield  {author} {\bibinfo {author} {\bibfnamefont {V.}~\bibnamefont
  {Kambersk{\'y}}},\ }\bibfield  {title} {\enquote {\bibinfo {title} {On the
  landau-lifshitz relaxation in ferromagnetic metals},}\ }\href@noop {}
  {\bibfield  {journal} {\bibinfo  {journal} {Canadian Journal of Physics}\
  }\textbf {\bibinfo {volume} {48}},\ \bibinfo {pages} {2906} (\bibinfo {year}
  {1970})}\BibitemShut {NoStop}%
\bibitem [{\citenamefont {Ingvarsson}\ \emph {et~al.}(1201)\citenamefont
  {Ingvarsson}, \citenamefont {Ritchie}, \citenamefont {Liu}, \citenamefont
  {Xiao}, \citenamefont {Slonczewski}, \citenamefont {Trouilloud},\ and\
  \citenamefont {Koch}}]{ingvarsson-scattering}%
  \BibitemOpen
  \bibfield  {author} {\bibinfo {author} {\bibfnamefont {S.}~\bibnamefont
  {Ingvarsson}}, \bibinfo {author} {\bibfnamefont {L.}~\bibnamefont {Ritchie}},
  \bibinfo {author} {\bibfnamefont {X.}~\bibnamefont {Liu}}, \bibinfo {author}
  {\bibfnamefont {G.}~\bibnamefont {Xiao}}, \bibinfo {author} {\bibfnamefont
  {J.}~\bibnamefont {Slonczewski}}, \bibinfo {author} {\bibfnamefont
  {P.}~\bibnamefont {Trouilloud}}, \ and\ \bibinfo {author} {\bibfnamefont
  {R.}~\bibnamefont {Koch}},\ }\bibfield  {title} {\enquote {\bibinfo {title}
  {Role of electron scattering in the magnetization relaxation of thin
  {N}i$_{81}${F}e$_{19}$ films},}\ }\href
  {http://dx.doi.org/10.1103/PhysRevB.66.214416} {\bibfield  {journal}
  {\bibinfo  {journal} {Phys. Rev, B, Condens, Matter Mater. Phys. (USA)}\
  }\textbf {\bibinfo {volume} {66}},\ \bibinfo {pages} {214416 -- 1} (\bibinfo
  {year} {2002/12/01})}\BibitemShut {NoStop}%
\bibitem [{\citenamefont {Silsbee}, \citenamefont {Janossy},\ and\
  \citenamefont {Monod}(1979)}]{silsbeePRB79}%
  \BibitemOpen
  \bibfield  {author} {\bibinfo {author} {\bibfnamefont {R.}~\bibnamefont
  {Silsbee}}, \bibinfo {author} {\bibfnamefont {A.}~\bibnamefont {Janossy}}, \
  and\ \bibinfo {author} {\bibfnamefont {P.}~\bibnamefont {Monod}},\ }\bibfield
   {title} {\enquote {\bibinfo {title} {Coupling between ferromagnetic and
  conduction-spin-resonance modes at a ferromagnetic normal-metal interface},}\
  }\href@noop {} {\bibfield  {journal} {\bibinfo  {journal} {Physical Review B
  (Condensed Matter)}\ }\textbf {\bibinfo {volume} {19}},\ \bibinfo {pages}
  {4382 -- 99} (\bibinfo {year} {1979})}\BibitemShut {NoStop}%
\bibitem [{\citenamefont {Tserkovnyak}, \citenamefont {Brataas},\ and\
  \citenamefont {Bauer}(2002)}]{tserkovPRL02sp}%
  \BibitemOpen
  \bibfield  {author} {\bibinfo {author} {\bibfnamefont {Y.}~\bibnamefont
  {Tserkovnyak}}, \bibinfo {author} {\bibfnamefont {A.}~\bibnamefont
  {Brataas}}, \ and\ \bibinfo {author} {\bibfnamefont {G.~E.~W.}\ \bibnamefont
  {Bauer}},\ }\bibfield  {title} {\enquote {\bibinfo {title} {Enhanced
  {G}ilbert damping in thin ferromagnetic films},}\ }\href@noop {} {\bibfield
  {journal} {\bibinfo  {journal} {Phys. Rev. Lett.}\ }\textbf {\bibinfo
  {volume} {88}},\ \bibinfo {pages} {117601} (\bibinfo {year}
  {2002})}\BibitemShut {NoStop}%
\bibitem [{\citenamefont {Tserkovnyak}\ \emph {et~al.}(2005)\citenamefont
  {Tserkovnyak}, \citenamefont {Brataas}, \citenamefont {Bauer},\ and\
  \citenamefont {Halperin}}]{tserk-rmp}%
  \BibitemOpen
  \bibfield  {author} {\bibinfo {author} {\bibfnamefont {Y.}~\bibnamefont
  {Tserkovnyak}}, \bibinfo {author} {\bibfnamefont {A.}~\bibnamefont
  {Brataas}}, \bibinfo {author} {\bibfnamefont {G.}~\bibnamefont {Bauer}}, \
  and\ \bibinfo {author} {\bibfnamefont {B.}~\bibnamefont {Halperin}},\
  }\bibfield  {title} {\enquote {\bibinfo {title} {Nonlocal magnetization
  dynamics in ferromagnetic heterostructures},}\ }\href@noop {} {\bibfield
  {journal} {\bibinfo  {journal} {Reviews in Modern Physics}\ }\textbf
  {\bibinfo {volume} {77}},\ \bibinfo {pages} {1375 -- 421} (\bibinfo {year}
  {2005})}\BibitemShut {NoStop}%
\bibitem [{\citenamefont {Zwierzycki}\ \emph {et~al.}(2005)\citenamefont
  {Zwierzycki}, \citenamefont {Tserkovnyak}, \citenamefont {Kelly},
  \citenamefont {Brataas},\ and\ \citenamefont {Bauer}}]{bauerPRB05}%
  \BibitemOpen
  \bibfield  {author} {\bibinfo {author} {\bibfnamefont {M.}~\bibnamefont
  {Zwierzycki}}, \bibinfo {author} {\bibfnamefont {Y.}~\bibnamefont
  {Tserkovnyak}}, \bibinfo {author} {\bibfnamefont {P.~J.}\ \bibnamefont
  {Kelly}}, \bibinfo {author} {\bibfnamefont {A.}~\bibnamefont {Brataas}}, \
  and\ \bibinfo {author} {\bibfnamefont {G.~E.~W.}\ \bibnamefont {Bauer}},\
  }\bibfield  {title} {\enquote {\bibinfo {title} {First-principles study of
  magnetization relaxation enhancement and spin transfer in thin magnetic
  films},}\ }\href {\doibase 10.1103/PhysRevB.71.064420} {\bibfield  {journal}
  {\bibinfo  {journal} {Phys. Rev. B}\ }\textbf {\bibinfo {volume} {71}},\
  \bibinfo {pages} {064420} (\bibinfo {year} {2005})}\BibitemShut {NoStop}%
\bibitem [{\citenamefont {Mizukami}, \citenamefont {Ando},\ and\ \citenamefont
  {Miyazaki}(2002{\natexlab{b}})}]{miz1}%
  \BibitemOpen
  \bibfield  {author} {\bibinfo {author} {\bibfnamefont {S.}~\bibnamefont
  {Mizukami}}, \bibinfo {author} {\bibfnamefont {Y.}~\bibnamefont {Ando}}, \
  and\ \bibinfo {author} {\bibfnamefont {T.}~\bibnamefont {Miyazaki}},\
  }\bibfield  {title} {\enquote {\bibinfo {title} {Magnetic relaxation of
  normal-metal {NM}/{N}i{F}e/{NM} films},}\ }\href@noop {} {\bibfield
  {journal} {\bibinfo  {journal} {Journal of Magnetism and Magnetic Materials}\
  }\textbf {\bibinfo {volume} {239}},\ \bibinfo {pages} {42 -- 4} (\bibinfo
  {year} {2002}{\natexlab{b}})}\BibitemShut {NoStop}%
\bibitem [{\citenamefont {Mizukami}, \citenamefont {Ando},\ and\ \citenamefont
  {Miyazaki}(2001)}]{miz2}%
  \BibitemOpen
  \bibfield  {author} {\bibinfo {author} {\bibfnamefont {S.}~\bibnamefont
  {Mizukami}}, \bibinfo {author} {\bibfnamefont {Y.}~\bibnamefont {Ando}}, \
  and\ \bibinfo {author} {\bibfnamefont {T.}~\bibnamefont {Miyazaki}},\
  }\bibfield  {title} {\enquote {\bibinfo {title} {The study on ferromagnetic
  resonance linewidth for {NM}/ 80{N}i{F}e/ {NM}, ({NM}={C}u, {T}a, {P}d and
  {P}t) films},}\ }\href@noop {} {\bibfield  {journal} {\bibinfo  {journal}
  {Japanese Journal of Applied Physics, Part 1 (Regular Papers, Short Notes and
  Review Papers)}\ }\textbf {\bibinfo {volume} {40}},\ \bibinfo {pages} {580 --
  5} (\bibinfo {year} {2001})}\BibitemShut {NoStop}%
\bibitem [{\citenamefont {Beaujour}\ \emph
  {et~al.}(2006{\natexlab{a}})\citenamefont {Beaujour}, \citenamefont {Chen},
  \citenamefont {Kent},\ and\ \citenamefont {Sun}}]{beaujourJAP06}%
  \BibitemOpen
  \bibfield  {author} {\bibinfo {author} {\bibfnamefont {J.}~\bibnamefont
  {Beaujour}}, \bibinfo {author} {\bibfnamefont {W.}~\bibnamefont {Chen}},
  \bibinfo {author} {\bibfnamefont {A.}~\bibnamefont {Kent}}, \ and\ \bibinfo
  {author} {\bibfnamefont {J.}~\bibnamefont {Sun}},\ }\bibfield  {title}
  {\enquote {\bibinfo {title} {Ferromagnetic resonance study of polycrystalline
  cobalt ultrathin films},}\ }\href@noop {} {\bibfield  {journal} {\bibinfo
  {journal} {Journal of Applied Physics}\ }\textbf {\bibinfo {volume} {99}},\
  \bibinfo {pages} {08N503} (\bibinfo {year} {2006}{\natexlab{a}})}\BibitemShut
  {NoStop}%
\bibitem [{\citenamefont {Beaujour}\ \emph
  {et~al.}(2006{\natexlab{b}})\citenamefont {Beaujour}, \citenamefont {Lee},
  \citenamefont {Kent}, \citenamefont {Krycka},\ and\ \citenamefont
  {Kao}}]{beaujourPRB06}%
  \BibitemOpen
  \bibfield  {author} {\bibinfo {author} {\bibfnamefont {J.-M.}\ \bibnamefont
  {Beaujour}}, \bibinfo {author} {\bibfnamefont {J.}~\bibnamefont {Lee}},
  \bibinfo {author} {\bibfnamefont {A.}~\bibnamefont {Kent}}, \bibinfo {author}
  {\bibfnamefont {K.}~\bibnamefont {Krycka}}, \ and\ \bibinfo {author}
  {\bibfnamefont {C.-C.}\ \bibnamefont {Kao}},\ }\bibfield  {title} {\enquote
  {\bibinfo {title} {Magnetization damping in ultrathin polycrystalline co
  films: evidence for nonlocal effects},}\ }\href@noop {} {\bibfield  {journal}
  {\bibinfo  {journal} {Physical Review B (Condensed Matter and Materials
  Physics)}\ }\textbf {\bibinfo {volume} {74}},\ \bibinfo {pages} {214405 -- 1}
  (\bibinfo {year} {2006}{\natexlab{b}})}\BibitemShut {NoStop}%
\bibitem [{\citenamefont {Lee}\ \emph {et~al.}(2008)\citenamefont {Lee},
  \citenamefont {Wen}, \citenamefont {Pathak}, \citenamefont {Janssen},
  \citenamefont {LeClair}, \citenamefont {Alexander}, \citenamefont {Mewes},\
  and\ \citenamefont {Mewes}}]{mewesJPD08}%
  \BibitemOpen
  \bibfield  {author} {\bibinfo {author} {\bibfnamefont {H.}~\bibnamefont
  {Lee}}, \bibinfo {author} {\bibfnamefont {L.}~\bibnamefont {Wen}}, \bibinfo
  {author} {\bibfnamefont {M.}~\bibnamefont {Pathak}}, \bibinfo {author}
  {\bibfnamefont {P.}~\bibnamefont {Janssen}}, \bibinfo {author} {\bibfnamefont
  {P.}~\bibnamefont {LeClair}}, \bibinfo {author} {\bibfnamefont
  {C.}~\bibnamefont {Alexander}}, \bibinfo {author} {\bibfnamefont
  {C.}~\bibnamefont {Mewes}}, \ and\ \bibinfo {author} {\bibfnamefont
  {T.}~\bibnamefont {Mewes}},\ }\bibfield  {title} {\enquote {\bibinfo {title}
  {Spin pumping in {C}o$_{56}${F}e$_{24}${B}$_{20}$ multilayer systems},}\
  }\href@noop {} {\bibfield  {journal} {\bibinfo  {journal} {Journal of Physics
  D: Applied Physics}\ }\textbf {\bibinfo {volume} {41}},\ \bibinfo {pages}
  {215001 (5 pp.) --} (\bibinfo {year} {2008})}\BibitemShut {NoStop}%
\bibitem [{\citenamefont {McMichael}\ and\ \citenamefont
  {Krivosik}(2004)}]{mcmichael-semiclassical}%
  \BibitemOpen
  \bibfield  {author} {\bibinfo {author} {\bibfnamefont {R.}~\bibnamefont
  {McMichael}}\ and\ \bibinfo {author} {\bibfnamefont {P.}~\bibnamefont
  {Krivosik}},\ }\bibfield  {title} {\enquote {\bibinfo {title} {Classical
  model of extrinsic ferromagnetic resonance linewidth in ultrathin films},}\
  }\href {http://dx.doi.org/10.1109/TMAG.2003.821564} {\bibfield  {journal}
  {\bibinfo  {journal} {IEEE Transactions on Magnetics}\ }\textbf {\bibinfo
  {volume} {40}},\ \bibinfo {pages} {2 -- 11} (\bibinfo {year}
  {2004})}\BibitemShut {NoStop}%
\bibitem [{Sch(1995)}]{Schreiber95SSC}%
  \BibitemOpen
  \bibfield  {title} {\enquote {\bibinfo {title} {Gilbert damping and g-factor
  in {F}e$_{x}${C}o$_{1-x}$ alloy films},}\ }\href@noop {} {\bibfield
  {journal} {\bibinfo  {journal} {Solid State Communications}\ }\textbf
  {\bibinfo {volume} {93}},\ \bibinfo {pages} {965 -- 968} (\bibinfo {year}
  {1995})}\BibitemShut {NoStop}%
\bibitem [{\citenamefont {Bass}\ and\ \citenamefont
  {Pratt}(2007)}]{bassJPCM07}%
  \BibitemOpen
  \bibfield  {author} {\bibinfo {author} {\bibfnamefont {J.}~\bibnamefont
  {Bass}}\ and\ \bibinfo {author} {\bibfnamefont {J.}~\bibnamefont {Pratt},
  \bibfnamefont {W.P.}},\ }\bibfield  {title} {\enquote {\bibinfo {title}
  {Spin-diffusion lengths in metals and alloys, and spin-flipping at
  metal/metal interfaces: an experimentalist's critical review},}\ }\href@noop
  {} {\bibfield  {journal} {\bibinfo  {journal} {Journal of Physics: Condensed
  Matter}\ }\textbf {\bibinfo {volume} {19}},\ \bibinfo {pages} {41 pp. --}
  (\bibinfo {year} {2007})}\BibitemShut {NoStop}%
\bibitem [{\citenamefont {Ahn}, \citenamefont {Shin},\ and\ \citenamefont
  {Pratt}(2008)}]{prattAPL08}%
  \BibitemOpen
  \bibfield  {author} {\bibinfo {author} {\bibfnamefont {C.}~\bibnamefont
  {Ahn}}, \bibinfo {author} {\bibfnamefont {K.-H.}\ \bibnamefont {Shin}}, \
  and\ \bibinfo {author} {\bibfnamefont {W.}~\bibnamefont {Pratt}},\ }\bibfield
   {title} {\enquote {\bibinfo {title} {Magnetotransport properties of
  {C}o{F}e{B} and {C}o/{R}u interfaces in the current-perpendicular-to-plane
  geometry},}\ }\href@noop {} {\bibfield  {journal} {\bibinfo  {journal}
  {Applied Physics Letters}\ }\textbf {\bibinfo {volume} {92}},\ \bibinfo
  {pages} {102509 -- 1} (\bibinfo {year} {2008})}\BibitemShut {NoStop}%
\end{thebibliography}

\printfigures

\end{document}